\documentclass[sigconf]{acmart}
\usepackage{textcomp, libertine}
\usepackage{booktabs} 
\usepackage[normalem]{ulem}
\usepackage{multirow}
\usepackage{bm}
\useunder{\uline}{\ul}{}

\usepackage{subcaption}
\usepackage{enumitem}
\usepackage[ruled]{algorithm2e}
\usepackage{dsfont}
\usepackage{pifont}
\hypersetup{
,bookmarksnumbered=true%
,hypertexnames=false%
,breaklinks=true%
,colorlinks=true%
,linkcolor=black%
,citecolor=black%
,urlcolor=black%
}

\definecolor{tealblue}{rgb}{0.21, 0.46, 0.53}
\definecolor{wildstrawberry}{rgb}{1.0, 0.26, 0.64}
\definecolor{ao(english)}{rgb}{0.0, 0.5, 0.0}

\def\ignore#1{}

\copyrightyear{2021}
\acmYear{2021}
\setcopyright{acmcopyright}\acmConference[SIGIR '21]{Proceedings of the 44th International ACM SIGIR Conference on Research and Development in Information Retrieval}{July 11--15, 2021}{Virtual Event, Canada}
\acmBooktitle{Proceedings of the 44th International ACM SIGIR Conference on Research and Development in Information Retrieval (SIGIR '21), July 11--15, 2021, Virtual Event, Canada}
\acmPrice{15.00}
\acmDOI{10.1145/3404835.3462987}
\acmISBN{978-1-4503-8037-9/21/07}

\begin{document}
\fancyhead{}


\title{Passage Retrieval for Outside-Knowledge \mbox{Visual Question Answering}}

\author{Chen Qu, Hamed Zamani, Liu Yang, W. Bruce Croft, Erik Learned-Miller}

\affiliation{%
	\institution{
		University of Massachusetts Amherst} 
	\streetaddress{140 Governors Dr.}
	\city{Amherst} 
	\state{MA} 
	\postcode{01003}
	\country{United States}
}
\email{{chenqu, zamani, lyang, croft, elm}@cs.umass.edu}

\begin{abstract}

In this work, we address multi-modal information needs that contain text questions and images by focusing on passage retrieval for outside-knowledge visual question answering. This task requires access to outside knowledge, which in our case we define to be a large unstructured passage collection. We first conduct sparse retrieval with BM25 and study expanding the question with object names and image captions. We verify that visual clues play an important role and captions tend to be more informative than object names in sparse retrieval. We then construct a dual-encoder dense retriever, with the query encoder being LXMERT~\cite{Tan2019LXMERTLC}, a multi-modal pre-trained transformer. We further show that dense retrieval significantly outperforms sparse retrieval that uses object expansion. Moreover, dense retrieval matches the performance of sparse retrieval that leverages human-generated captions.

\end{abstract}

\keywords{Dense Retrieval; Multi-Modal; Visual Question Answering}

\ccsdesc[500]{Information systems~Question answering}
\ccsdesc[500]{Information systems~Multimedia and multimodal retrieval}

\maketitle

\section{Introduction}
\label{sec:intro}
\begin{figure}[t]
    \centering
    \includegraphics[width=0.47\textwidth]{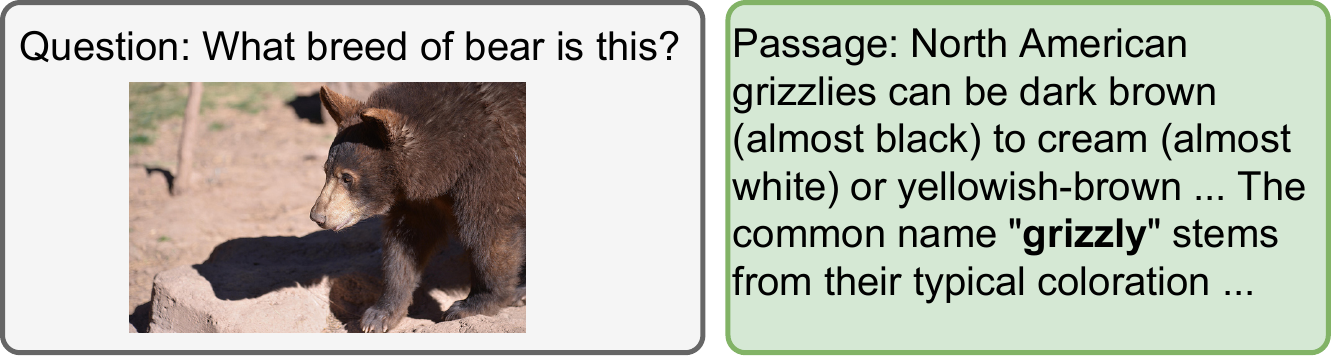}
    \vspace{-0.4cm}
    \caption{An example of passage retrieval for OK-VQA. Boldface denotes a potential answer. Image \copyright gsloan, \url{https://www.flickr.com/photos/gsloan/8137199999/}}
    \label{fig:examples}
    \vspace{-0.4cm}
\end{figure} 

Recent work on Question Answering (QA)~\cite{squad,iart,orconvqa} mostly focuses on uni-modal information needs, i.e., text- or voice-based questions (voice input can be considered as text after automatic transcription).
However, many information needs, such as the one in Fig.~\ref{fig:examples}, would be inconvenient or hard to explain without a picture. 
This motivates the study of methods that can handle multi-modal information needs containing both text questions and images~\cite{Deldjoo:2021,Lien:2020}.

Specifically, we focus on a Visual QA (VQA) task referred to as Outside-Knowledge VQA (OK-VQA). Classic VQA benchmarks~\cite{Malinowski2014AMA,Yu2015VisualMF,Zhu2016Visual7WGQ,mscoco,Agrawal2015VQAVQ} and models~\cite{Benyounes2017MUTANMT,Kim2018BilinearAN,Malinowski2015AskYN,Xiong2016DynamicMN,Lu2016HierarchicalQC,Fukui2016MultimodalCB} mainly focus on questions about counting, visual attributes, or other visual detection tasks, whose answers can be found in the given image. In contrast, images in our task help to define the information need, instead of simply being the knowledge source by which the question is answered. OK-VQA resembles open-domain QA~\cite{trec8} in the sense that both tasks require access to an outside and open knowledge resource, e.g., a large collection of passages, to answer the questions. Open-domain QA systems typically follow a retrieve-and-read paradigm~\cite{Karpukhin2020DensePR,orqa,orconvqa,drqa,Qu2020RocketQAAO,Xiong2020AnsweringCO,Qu2021WeaklySupervisedOC}, where the system first retrieves a number of documents (passages) from a collection and then extracts answers from them. This paradigm is less studied for multi-modal information needs, which is the focus of this paper. In this work, we focus on the retrieval phase for OK-VQA as illustrated in Fig.~\ref{fig:examples}. 

Unlike previous knowledge-based VQA work that retrieves knowledge from a knowledge base~ \cite{Gardres2020ConceptBertCR,Narasimhan2018OutOT,Narasimhan2018StraightTT,Wang2017ExplicitKR,Wang2018FVQAFV,Wu2016AskMA,Li2017IncorporatingEK,Yu2020CrossmodalKR,Zhu2020MuckoMC} or using a Wikipedia Search API~\cite{okvqa}, we systematically study passage retrieval for OK-VQA with \textit{generic} information retrieval approaches
so that our methods can be applied to a wider range of \emph{unstructured} knowledge resources.
In particular, we seek answers to the following research questions: (\textbf{RQ1}) How helpful are the visual signals in OK-VQA? (\textbf{RQ2}) What is the most effective way to incorporate visual signals into sparse retrieval models that are based on term matching? (\textbf{RQ3}) How well does dense retrieval~\cite{Luan2020SparseDA,Karpukhin2020DensePR,Xiong2020ApproximateNN,orqa,Guu2020REALMRL} work with multi-modal information needs? 

To answer these important research questions, we study passage retrieval for OK-VQA queries with a large Wikipedia passage collection. We first conduct sparse retrieval with BM25. We investigate the performance of expanding the original question with different human-annotated object names and image captions. We further study the impact of using different rank fusion methods for different expansion types. We verify that visual clues play an important role in our task. In particular, captions tend to be more informative than object names in sparse retrieval. We further reveal that it is desirable to exploit the most salient matching signal (CombMAX~\cite{Lee1997AnalysesOM,Fox1993CombinationOM}) when using object expansion
while it is better to consider the matching signals for all captions with CombSUM~\cite{Lee1997AnalysesOM,Fox1993CombinationOM} or Reciprocal Rank Fusion~\cite{Hu2018ReinforcedMR} when we expand with human-generated captions.

We then adopt a dual-encoder architecture to construct a learnable dense retriever following previous work~\cite{Luan2020SparseDA,Karpukhin2020DensePR,Xiong2020ApproximateNN,orconvqa,orqa}. We employ LXMERT~\cite{Tan2019LXMERTLC}, a pre-trained Transformer model~\cite{transformer}, as our multi-modal query encoder to encode both the text question and the image as an information need. We observe that our dense retriever achieves a statistically significant performance improvement over sparse retrieval that leverages object expansion, demonstrating the effectiveness of dense retrieval with a multi-modal query encoder. Furthermore, our dense retriever manages to match the performance of sparse retrieval with caption expansion, even though the latter leverages human-generated captions that are often highly informative. 
Our research is one of fundamental steps for future studies on retrieval-based OK-VQA. 
Our code is released for research purposes.\footnote{\url{https://github.com/prdwb/okvqa-release}}
\section{Passage Retrieval for OK-VQA}
\label{sec:our-approach}

\subsection{Task Definition}
\label{subsec:task}
We are given an information need (query) denoted as $Q_i= \langle q_i, v_i \rangle$. It consists of a text question $q_i$ and an image $v_i$. The task is to retrieve $k$ passages 
that can be used to fulfill $Q_i$, from a large passage collection. Following the work on open-domain QA~\cite{orqa,orconvqa,Karpukhin2020DensePR}, a passage is deemed as relevant if it contains the ground truth answer.

\subsection{Sparse Retrieval}
\label{subsec:sparse}

The backbone of our sparse retrieval approach is BM25,
which works with text queries. Therefore, we expand $q_i$ with different textual descriptions of visual clues to construct the BM25 queries. Visual signals in an image are typically expressed in two forms. The first form is a set of object names $\{o_i^1, o_i^2, \cdots \}$ produced by an object detector. Each object is a Region of Interest (RoI) that reveals a meaningful component of the image.
The second form is a set of captions $\{c_i^1, c_i^2, \cdots\}$ produced by an image descriptor to describe the image. 
We adopt human-annotated object names and captions for sparse retrieval. Although the human annotations do not necessarily give the performance upper bound, they would be strong baselines for dense retrieval and make sure that our analysis will not be affected by the quality of automatic annotations produced by object detectors and image descriptors. 
We study different expansion of visual signals as follows:
\begin{itemize}[leftmargin=*, noitemsep, topsep=0pt]
    \item \textbf{BM25-Orig}: taking the original $q_i$ only, i.e., $Q_i^\text{orig}=\{q_i\}$.
    \item \textbf{BM25-Obj} (object expansion): appending each one of the object names to $q_i$, i.e., $Q_i^\text{obj}=\{q_i + o_i^1, q_i + o_i^2, \cdots \}$. 
    \item \textbf{BM25-Cap} (caption expansion): appending each one of the captions to $q_i$, i.e., $Q_i^\text{cap}=\{q_i + c_i^1, q_i + c_i^2, \cdots \}$. 
    \item \textbf{BM25-All}: taking the union of the above queries
    , i.e., $Q_i^\text{all}=Q_i^\text{orig} \cup Q_i^\text{obj} \cup Q_i^\text{cap}$.
\end{itemize}
Since $Q_i^\text{obj/cap/all}$ contains multiple BM25 queries for the same information need $Q_i$, we need rank fusion methods to consolidate the ranked lists $R$ generated by queries within each query set.
This resembles an ensemble process to combine results obtained with different visual signals. 
We consider \textbf{CombMAX}~\cite{Lee1997AnalysesOM,Fox1993CombinationOM} (taking the maximum score of a passage in different ranked lists), \textbf{CombSUM}~\cite{Lee1997AnalysesOM,Fox1993CombinationOM} (taking the sum of scores of a passage in different ranked lists), and \textbf{RRF} (Reciprocal Rank Fusion)~\cite{Cormack2009ReciprocalRF} (the fusion score for a passage $p$ is defined as $\sum_{r \in R} \frac{1}{const + r(p)}$, where $r(\cdot)$ is the rank of $p$).
CombMAX could help the model be more robust to distracting visual signals while the other two fusion approaches make sure that the impacts of lower-ranked passages do not vanish.

\subsection{Dense Retrieval}
\label{subsec:methods}
\begin{figure}[t]
    \centering
    \includegraphics[width=0.47\textwidth]{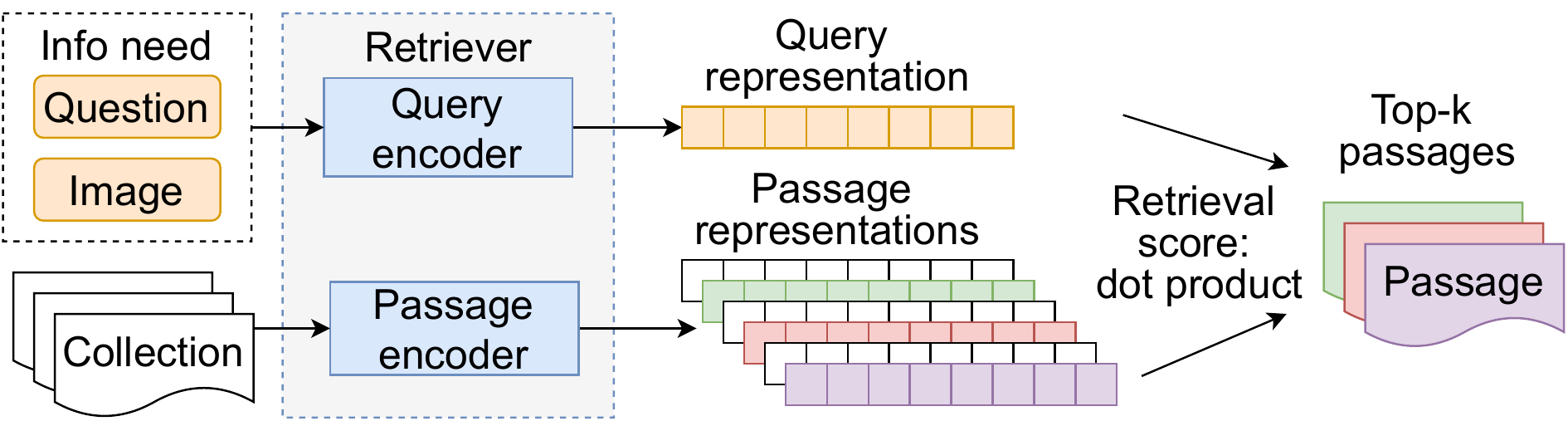}
    \vspace{-0.4cm}
    \caption{Dense retrieval with neural dual encoders.}
    \label{fig:dense}
\end{figure} 

Following previous work~\cite{Karpukhin2020DensePR,Xiong2020ApproximateNN,orconvqa,orqa}, we adopt a dual-encoder architecture to construct a learnable retriever. The retrieval process is ``dense'' in the sense that the queries and passages are encoded to low-dimensional dense vectors, as opposed to the high-dimensional sparse vectors used in sparse retrieval. As shown in Fig.~\ref{fig:dense}, the retriever consists of a query encoder and a passage encoder.

\subsubsection{\textbf{Query Encoder}}
We adopt the LXMERT model~\cite{Tan2019LXMERTLC} as the query encoder since it can encode both the question and image components of $Q_i$. LXMERT is a pre-trained Transformer model~\cite{transformer} designed to learn vision and language connections. It consists of three encoders, an object relationship encoder, a language encoder, and a cross-modality encoder. The first two single-modality encoders function in a similar way to BERT~\cite{bert} except that the object relationship encoder works with a set of object detections produced by a Faster R-CNN~\cite{Ren2015FasterRT} model pre-trained on Visual Genome~\cite{Anderson2018BottomUpAT,Krishna2016VisualGC}. Each detection representation can be considered as an ``image token embedding'' that consists of its RoI features (fixed) and position features (trainable). The cross-modality encoder conducts bi-directional cross attention between vision and language representations. 
We refer our readers to the LXMERT paper~\cite{Tan2019LXMERTLC} for further details. We project the cross-modality output of LXMERT to an $n$-dimensional query representation. The dense retriever with a LXMERT query encoder is referred to as \textbf{Dense-LXMERT}. To adopt a deeper analytical view, we also consider BERT as the query encoder, which only works with the question component of the query, resulting in the \textbf{Dense-BERT} model. 

\subsubsection{\textbf{Passage Encoder}}
We use BERT as the passage encoder and project the \texttt{[CLS]} representation to an $n$-dimensional passage representation. The retrieval score is defined as the dot product of the query and passage representations.
After training, we encode all passages in the collection during an offline process. At inference time, we use Faiss~\cite{faiss}
for maximum inner product search. 

\subsubsection{\textbf{Training}}
\label{subsubsec:training}

We train the dual-encoder retriever with a set of training instances. Each instance is denoted as $\langle Q_i, p_i^+, p_i^- \rangle$, where $p_i^+$ is a positive passage that contains the answer while $p_i^-$ is a negative passage that does not contain the answer. In our case, we select $p_i^-$ from the top passages retrieved by a sparse retrieval method. Thus, $p_i^-$ can be referred to as a \textit{retrieved negative}. We present more details on constructing the training data in Sec.~\ref{subsubsec:dense-data}. 
In addition to the retrieved negatives, one might want to take advantage of the other passages in the batch as in-batch negatives. Although in-batch negatives resemble randomly sampled negatives that can be less effective~\cite{Karpukhin2020DensePR}, it is extremely efficient since passage representations can be reused within the batch. \citet{Karpukhin2020DensePR} studied 
combining in-batch negatives with retrieved negatives for uni-modal queries. 
We further dig into this topic for multi-modal queries. We consider the following negative sampling strategies:
\begin{itemize}[leftmargin=*, noitemsep, topsep=0pt]
    \item \textbf{R-Neg}
    : using the \textbf{r}etrieved \textbf{neg}ative passage only.
    
    \item \textbf{R-Neg+IB-Neg}
    : using the \textbf{r}etrieved \textbf{neg}ative, along with all other \textbf{i}n-\textbf{b}atch \textbf{neg}ative passages of other instances. 
    
    \item \textbf{R-Neg+IB-Pos}
    : using the \textbf{r}etrieved \textbf{neg}ative, along with all other \textbf{i}n-\textbf{b}atch \textbf{pos}itive passages of other instances.  
    
    \item \textbf{R-Neg+IB-All}
    : using the \textbf{r}etrieved \textbf{neg}ative, along with \textbf{all} other \textbf{i}n-\textbf{b}atch passages, except for $p_i^+$.
    The same query can be paired with different positive and negative passages to augment the training data as suggested in Sec.~\ref{subsubsec:dense-data}. Therefore, the queries within a batch can coincide even with random batching. In this case, the misuse of a positive passage as negative may hinder the learning process. We empirically examine whether this concern holds by comparing R-Neg+IB-All/Pos to R-Neg+IB-Neg.
\end{itemize}

Following previous work~\cite{orconvqa,Karpukhin2020DensePR}, we use cross entropy loss to maximize the probability of the positive passage given the negatives identified above. We then average the losses for queries in the batch.


\section{Experiments}
\label{sec:exp}
\subsection{Experimental Setup}
\label{subsec:setup}
\subsubsection{\textbf{Dataset}}
\label{subsubsec:data}
Our retrieval dataset is based on the OK-VQA dataset~\cite{okvqa}, where all questions require outside knowledge.\footnote{\url{https://okvqa.allenai.org/index.html}}
The images in the OK-VQA dataset are from the COCO dataset~\cite{mscoco}.
We take the original training queries as our training queries and split the original validation queries into our validation and testing queries. In terms of the collection, we take the Wikipedia passage collection with 11 million passages created by previous work \cite{orconvqa}.\footnote{\url{https://ciir.cs.umass.edu/downloads/ORConvQA/all_blocks.txt.gz}}
Each passage contains at most 384 ``wordpieces''~\cite{bert} with intact sentence boundaries. Data statistics are presented in Tab.~\ref{tab:stats}.

\subsubsection{\textbf{Data Construction for Dense Retrieval}}
\label{subsubsec:dense-data}

We create the training instances described in Sec.~\ref{subsubsec:training} using retrieved passages of sparse retrieval (see configuration details in Sec.~\ref{subsubsec:details}).
A passage is identified to be positive if it contains an exact match (case-insensitive) of a ground truth answer. The other retrieved passages are considered as negatives. We take the top 5 positive passages, each repeated 5 times (for augmentation), to construct training instances with random retrieved negatives.
In addition, we put together a small validation collection by taking the top 20 passages for each question. Data statistics are presented in Tab.~\ref{tab:stats}.

\setlength{\tabcolsep}{2.5pt}
\begin{table}[t]
\caption{Data statistics.}
\label{tab:stats}
\vspace{-0.4cm}
\footnotesize
\begin{tabular}{@{}ccccc@{}}
\toprule
Split & \#. questions & \#. BM25 queries & \#. training instances & \#. passages in collection  \\ \midrule
Train & 9,009         & 81,100      & 211,200       & N/A                        \\
Val   & 2,523         & 22,352      & N/A           & 34,059                     \\
Test  & 2,523         & 22,573      & N/A           & 11,000,000                 \\ \bottomrule
\end{tabular}
\end{table}


\subsubsection{\textbf{Evaluation Metrics}}
\label{subsubsec:metrics}

We focus on passage retrieval for multi-modal information needs as the first step in the OK-VQA pipeline. The output of the retrieval process will be used by a reader model to extract the answer. Therefore, following previous work~\cite{Karpukhin2020DensePR,drqa}, we use precision-oriented metrics to evaluate the performance of our models. Precisely, we use Mean Reciprocal Rank and Precision with the ranking cut-off of 5 (MRR@5 and P@5) as our metrics.

\subsubsection{\textbf{Implementation Details}}
\label{subsubsec:details}
For sparse retrieval, we use BM25 in Anserini (v0.5.1).\footnote{\url{https://github.com/castorini/anserini}} We tune $k_1 \in [0.5, 1.5]$ and $b \in [0.2, 0.8]$ with steps of 0.2 based on validation MRR. The best setting is $k_1=1.1,b=0.4$. The constant in RRF is set to 60 \cite{Cormack2009ReciprocalRF, Benham2017RiskRewardTI}. Human-annotated object names and captions are from the COCO dataset~\cite{mscoco}.
For dense retrieval, we use the HuggingFace transformers library\footnote{\url{https://github.com/huggingface/transformers}} for the implementations of LXMERT and BERT. We set the maximum sequence length of the query encoder to 20~\cite{Tan2019LXMERTLC}, that of the passage encoder to 384, the projection size ($n$) for the query/passage representations to 768, the learning rate to 1e-5, the batch size to 4 per GPU, and the number of fine-tuning epochs to 2. We adopt R-Neg+IB-All as the negative sampling strategy. We save checkpoints every 5,000 steps and evaluate on the validation set to select the best model for the test set. The training time is 10 hours for Dense-BERT and 12 hours for Dense-LXMERT. All models are trained with 4 GPUs with mixed-precision training. Warm-up takes 10\% of the total steps. The training instances are constructed with the top 100 retrieved passages for each question using BM25-Cap (CombSUM) with the default BM25 configuration in Anserini ($k_1=0.9,b=0.4$).

\subsection{Main Results}
\label{subsec:main-results}

\subsubsection{\textbf{Sparse Retrieval}}
\label{subsubsec:sparse-results}
We present the results for sparse retrieval in Tab.~\ref{tab:sparse-results} to answer \textbf{RQ1} and \textbf{RQ2} raised in Sec.~\ref{sec:intro}. First, we observe that approaches that consider visual signals outperform BM25-Orig by a large margin, verifying that visual clues are helpful in our task. We then compare BM25 with different forms of visual clues. Methods with captions (BM25-Cap/All) outperform object expansion, indicating that captions are more informative than object names. This makes sense since captions typically cover important objects descriptively. BM25-All does not benefit from incorporating both objects and captions. On the contrary, objects can be distracting and hurt the performance gain from captions. Finally, we compare different rank fusion methods. When objects are being considered (BM25-Obj/All), CombMAX yields the best performance since it is robust to potentially misleading objects by only considering objects with the best matching score. On the other hand, CombSUM and RRF work well with caption expansion. Their ability to consider the impact of all captions is desirable since captions are closely connected to the image and can be diverse and complementary. The best performing approach is BM25-Cap with CombSUM.

\setlength{\tabcolsep}{3pt}
\begin{table}[t]
\caption{Sparse retrieval results. Boldface denotes the best performance within each group and underscores denote the best overall results. $\blacktriangle i$ denotes that the gain with respect to the best method in group $i$ has statistically significance with $p < 0.05$ tested by the Student’s paired t-test.}
\label{tab:sparse-results}
\footnotesize
\vspace{-0.4cm}
\begin{tabular}{l|c|ll|ll}
\toprule
\multicolumn{2}{c|}{Methods}             & \multicolumn{2}{c|}{Val}                      & \multicolumn{2}{c}{Test}                      \\ \midrule
\multicolumn{1}{c|}{Expansion} & Fusion  & MRR@5                 & P@5                   & MRR@5                 & P@5                   \\ \midrule
1. BM25-Orig                   & N/A     & 0.2565                & 0.1772                & 0.2637                & 0.1755                \\ \midrule
\multirow{3}{*}{2. BM25-Obj}   & CombMAX & \textbf{0.3772}$^{\blacktriangle 1}$       & \textbf{0.2667}$^{\blacktriangle 1}$       & \textbf{0.3686}$^{\blacktriangle 1}$       & \textbf{0.2541}$^{\blacktriangle 1}$       \\
                               & CombSUM & 0.3493                & 0.2395                & 0.3406                & 0.2322                \\
                               & RRF     & 0.3389                & 0.2291                & 0.3292                & 0.2213                \\ \midrule
\multirow{3}{*}{3. BM25-Cap}   & CombMAX & 0.4547                & 0.3294                & 0.4534                & 0.3230                \\
                               & CombSUM & {\ul \textbf{0.4727}}$^{\blacktriangle 1,2,4}$ & {\ul \textbf{0.3483}}$^{\blacktriangle 1,2,4}$ & {\ul \textbf{0.4622}}$^{\blacktriangle 1,2}$ & {\ul \textbf{0.3367}}$^{\blacktriangle 1,2,4}$ \\
                               & RRF     & 0.4689                & 0.3440                & 0.4585                & 0.3346                \\ \midrule
\multirow{3}{*}{4. BM25-All}   & CombMAX & \textbf{0.4550}$^{\blacktriangle 1,2}$      & \textbf{0.3293}$^{\blacktriangle 1,2}$        & \textbf{0.4533}$^{\blacktriangle 1,2}$        & \textbf{0.3233}$^{\blacktriangle 1,2}$        \\
                               & CombSUM & 0.4490                & 0.3241                & 0.4396                & 0.3126                \\
                               & RRF     & 0.4322                & 0.3069                & 0.4260                & 0.2956                \\ \bottomrule
\end{tabular}
\end{table}

\subsubsection{\textbf{Dense Retrieval}}
\label{subsubsec:dense-results}
We present the dense retrieval results, along with the best sparse retrieval results in Tab.~\ref{tab:dense-results} to answer \textbf{RQ3}. We first compare retrieval without visual clues: we observe that Dense-BERT outperforms BM25-Orig by a large margin, verifying the capability of dense retrieval. We further explain that this capability is contingent upon the negative sampling strategy used during training in Sec.~\ref{subsubsec:negative-sampling-analysis}. Moreover, Dense-BERT even surpasses BM25-Obj that considers visual signals. 
This could be due to the tendency of Dense-BERT to retrieve passages containing frequent answers.
We speculate this kind of overfitting is caused by the lack of visual signals. Further analysis can be found in Sec.~\ref{subsec:additional-results}.

We further observe that Dense-LXMERT significantly outperforms BM25-Obj. 
Dense-LXMERT leverages both the RoI features and position features in object detection to learn object relations, which can be more effective than using ground truth object names in sparse retrieval.
We then compare Dense-LXMERT with BM25-Cap/All. These sparse retrieval methods consider human-annotated image captions that are highly informative and descriptive. On the contrary, Dense-LXMERT has to learn the importance of the objects and the relation among them with object-level features. In this unfavorable situation, Dense-LXMERT still manages to match the performance of BM25-Cap/All. Although BM25-Cap is slightly better, the margins are statistically \textit{insignificant}. 
Finally, we observe that Dense-LXMERT significantly outperforms Dense-BERT, further validating the use of a multi-modal query encoder. 

\setlength{\tabcolsep}{1pt}
\begin{table}[t]
\caption{Dense retrieval results.
Refer to Tab.~\ref{tab:sparse-results} for notations. $\triangle i$ denotes the statistical significance is obtained with $0.05 < p < 0.1$.
Note that \textit{BM25-Obj/Cap/All has access to the ground truth object names and captions. 
}
}
\label{tab:dense-results}
\footnotesize
\vspace{-0.4cm}
\begin{tabular}{l|l|ll|ll}
\toprule
\multicolumn{2}{c|}{\multirow{2}{*}{Methods}}   & \multicolumn{2}{c|}{Val}                             & \multicolumn{2}{c}{Test}                            \\ \cmidrule{3-6} 
\multicolumn{2}{c|}{}                           & \multicolumn{1}{l}{MRR@5} & \multicolumn{1}{l|}{P@5} & \multicolumn{1}{l}{MRR@5} & \multicolumn{1}{l}{P@5} \\ \midrule
\multirow{4}{*}{\rotatebox{90}{Sparse}} & 1. BM25-Orig          & 0.2565                    & 0.1772                   & 0.2637                    & 0.1755                  \\
                        & 2. BM25-Obj (CombMAX) & 0.3772$^{\blacktriangle 1}$                    & 0.2667$^{\blacktriangle 1}$                   & 0.3686$^{\blacktriangle 1}$                    & 0.2541$^{\blacktriangle 1}$                  \\
                        & 3. BM25-Cap (CombSUM) & {\ul \textbf{0.4727}}$^{\blacktriangle 1,2,4 \ \triangle 5}$     & {\ul \textbf{0.3483}}$^{\blacktriangle 1,2,4,5}$    & {\ul \textbf{0.4622}}$^{\blacktriangle 1,2,5}$     & {\ul \textbf{0.3367}}$^{\blacktriangle 1,2,4,5}$   \\
                        & 4. BM25-All (CombMAX) & 0.4550$^{\blacktriangle 1,2}$                    & 0.3293$^{\blacktriangle 1,2 \ \triangle 5}$                   & 0.4533$^{\blacktriangle 1,2 \ \triangle 5}$                    & 0.3233$^{\blacktriangle 1,2,5}$                  \\ \midrule
\multirow{2}{*}{\rotatebox{90}{Dense}}  & 5. Dense-BERT         & 0.4555$^{\blacktriangle 1,2}$                    & 0.3155$^{\blacktriangle 1,2}$                   & 0.4325$^{\blacktriangle 1,2}$                    & 0.3058$^{\blacktriangle 1,2}$                  \\
                        & 6. Dense-LXMERT       & \textbf{0.4704}$^{\blacktriangle 1,2 \ \triangle 5}$           & \textbf{0.3364}$^{\blacktriangle 1,2,5}$          & \textbf{0.4526}$^{\blacktriangle 1,2,5}$           & \textbf{0.3329}$^{\blacktriangle 1,2,5}$         \\ \bottomrule
\end{tabular}\end{table}

\subsection{Additional Results}
\label{subsec:additional-results}
We provide additional analysis to study the impact of the projection size and negative sampling strategies with validation performance.

\subsubsection{\textbf{Impact of projection size}}
We present the impact of the dimensionality $n$ of query/passage representations in Fig.~\ref{fig:proj_size_analysis}. We observe that a larger projection size always leads to better performance, although the performance gain seems to be insignificant for LXMERT after $n=256$. We set $n=768$ as reported in Sec.\ref{subsubsec:details} since it gives the best performance for both Dense-LXMERT and Dense-BERT. When working with a much larger collection than ours (ours has 11 million passages), one might want to use $n=256$ since it offers similar performance with less memory consumption.

\subsubsection{\textbf{Impact of negative sampling}}
\label{subsubsec:negative-sampling-analysis}
\begin{figure}[t]
    \centering
    \begin{subfigure}[b]{0.24\textwidth}
        \includegraphics[width=\textwidth]{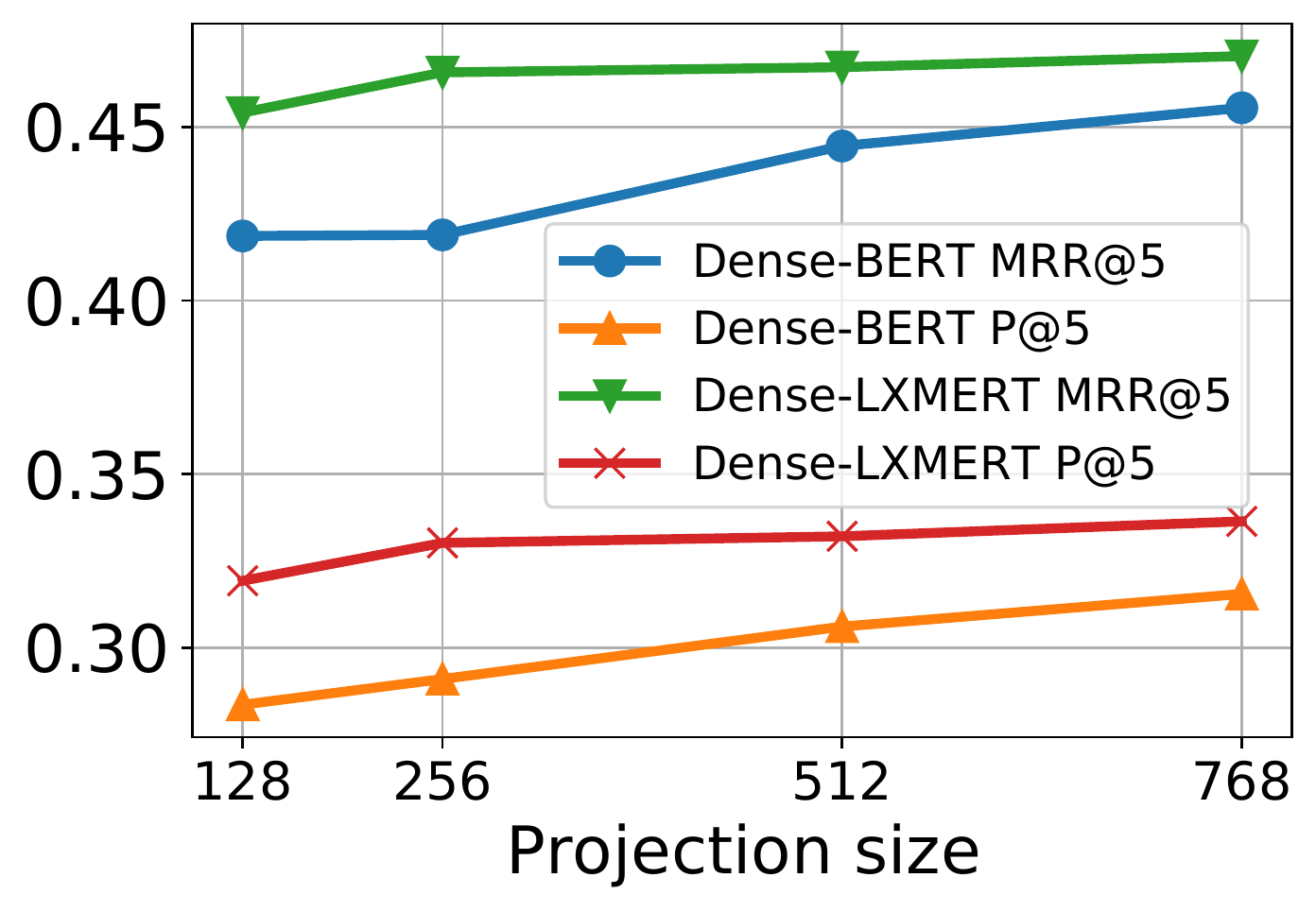}
        \vspace{-0.5cm}
        \caption{Impact of projection size $n$.}
        \label{fig:proj_size_analysis}
    \end{subfigure}
    \begin{subfigure}[b]{0.23\textwidth}
        \includegraphics[width=\textwidth]{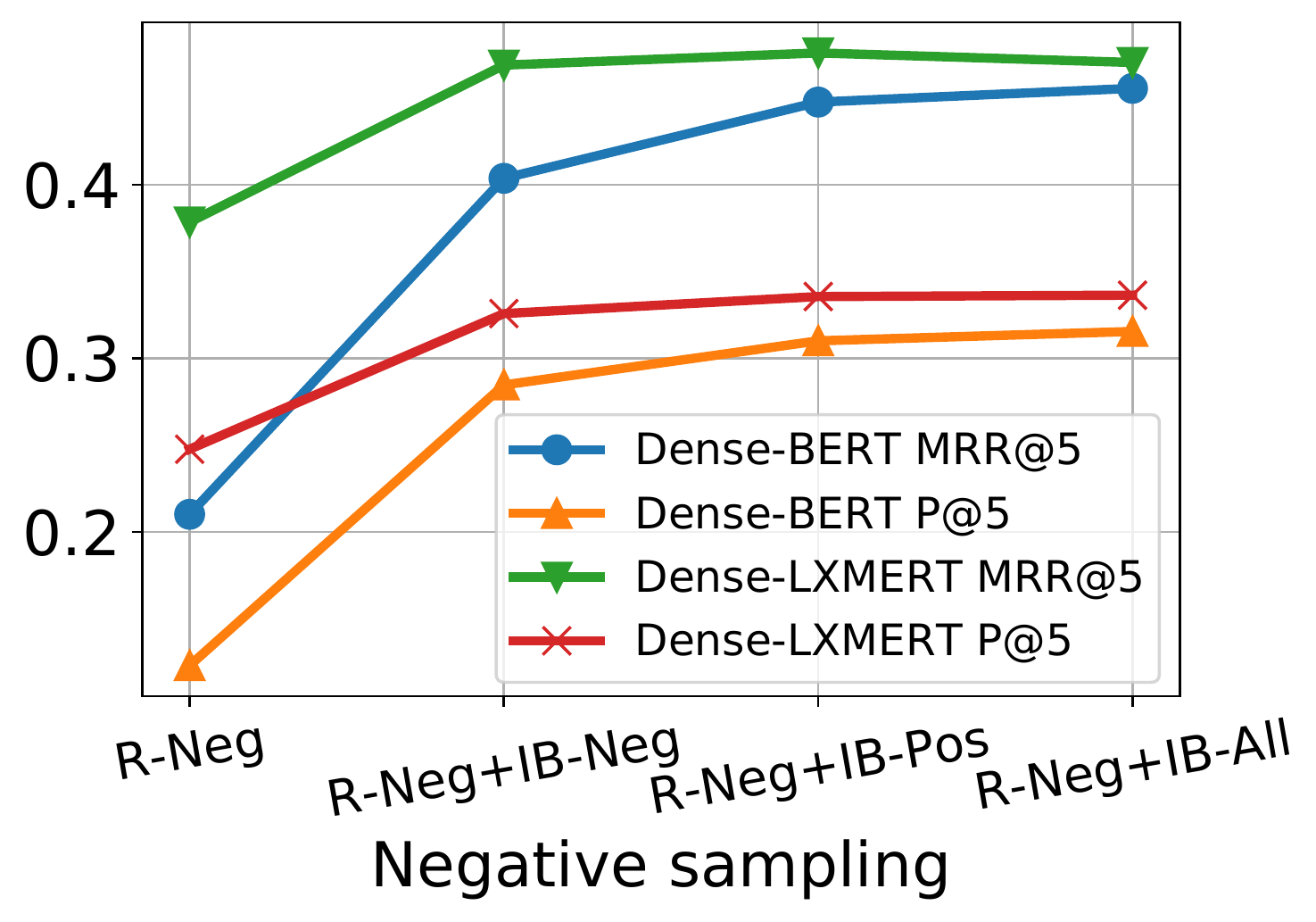}
        \vspace{-0.5cm}
        \caption{Impact of negative sampling.}
        \label{fig:negative_sampling_analysis}
    \end{subfigure}
    \vspace{-0.5cm}
    \caption{Additional results.}
    \label{fig:history}
\end{figure}

The desirable performance of dense retrieval is contingent upon the negative sampling strategy. We present the impact of different sampling methods described in Sec.~\ref{subsubsec:training} in Fig.~\ref{fig:negative_sampling_analysis}. We observe that combining retrieved negatives with in-batch negatives dramatically improves the model performance, verifying the observations in \citet{Karpukhin2020DensePR} for multi-modal queries.
Also, different choices of in-batch negatives (R-Neg+IB-Neg/Pos/All) give a similar performance for LXMERT, indicating that coinciding questions in the same batch should not be a concern for our batch size and data size reported in Sec.~\ref{subsec:setup}.

Both analyses show that Dense-BERT is more demanding on larger model capacity (larger projection size) and more negative samples. We speculate that BERT is overfitting the patterns in the training data since it lacks important visual clues for matching. In comparison, Dense-LXMERT is less sensitive to reasonably-chosen projection sizes and negative sampling strategies because it can learn matching signals from both language and vision clues. 

\section{Conclusions and Future Work} 
\label{sec:conclusion}
We study passage retrieval for OK-VQA with sparse and dense retrieval and verify visual clues play an important role. We discover that captions are more informative than object names in sparse retrieval and CombMAX works well with object expansion while CombSUM and RRF are better for caption expansion. 
We further show a dense retriever with a multi-modal query encoder can significantly outperform sparse retrieval with object expansion and even matches the performance of that with human-generated captions. 
In the future, we will consider using automatic captions for sparse retrieval and study answer extraction to complete the QA pipeline.

\begin{acks}
This work was supported in part by the Center for Intelligent Information Retrieval. Any opinions, findings and conclusions or recommendations expressed in this material are those of the authors and do not necessarily reflect those of the sponsors.
\end{acks}

\newpage
\bibliographystyle{abbrvnat}
\balance 
\bibliography{acmart} 

\end{document}